\documentclass[aps,preprint,amsmath,amssymb,t1,11pt]{revtex4}
\usepackage{graphicx}
\usepackage{epstopdf}
\usepackage{xcolor}
\usepackage{slashed}

\begin{document}
\draft

\title{Search for the electromagnetic properties of the neutrinos at the
HL-LHC and the FCC-hh}

\author{M. K{\"o}ksal}
\email[]{mkoksal@cumhuriyet.edu.tr}
\affiliation{Department of Physics, Sivas Cumhuriyet University, 58140, Sivas, T\"{u}rkiye.}

\author{A. Senol}
\email[]{senol_a@ibu.edu.tr}
\affiliation{Department of Physics, Bolu Abant Izzet Baysal University, 14280, Bolu, T\"{u}rkiye.}

\author{H. Denizli}
\email[]{denizli_h@ibu.edu.tr}
\affiliation{Department of Physics, Bolu Abant Izzet Baysal University, 14280, Bolu T\"{u}rkiye.}

\date{\today}

\begin{abstract}

The $\nu\bar{\nu}\gamma \gamma$ couplings parametrized with the non-standard dimension-seven operators defined by the Effective Field Theory framework are investigated through the process $pp\to \nu\bar{\nu}\gamma$  at the High Luminosity-LHC and the Future Circular proton-proton Collider. The effective Lagrangian of $\nu\bar{\nu}\gamma \gamma$ couplings is implemented into FeynRules to generate a UFO module inserted into Madgraph to generate both background and signal events. These events are then passed through Pythia 8 for parton showering and Delphes to include realistic detector effects. The sensitivities on $\nu\bar{\nu}\gamma \gamma$ couplings are obtained at $95\%$ confidence level. We show that the analysis of the signal emerging from the process $pp\to \nu\bar{\nu}\gamma$ allows to improve constraints on $\nu\bar{\nu}\gamma \gamma$ couplings given by the LEP collaboration.
\end{abstract}


\vspace{5mm}

\maketitle


\section{Introduction}

The new phase of particle physics with advanced machine configurations at the LHC and beyond aims to enhance precision by increasing center-of-mass energy and luminosity in precision measurements and the search for new physics. The High Luminosity-LHC (HL-LHC) upgrade project will provide an integrated luminosity of 3000 fb$^{-1}$ at center-of-mass energy of 14 TeV  over 12 years with an annual rate of 250 fb$^{-1}$ \cite{lhc,yel}. However, the European Strategy Group's Update of the European Strategy for Particle Physics recommends exploring the feasibility of a future circular collider at CERN with a center-of-mass energy of 100 TeV with an electron-positron Higgs and electro-weak factory as a possible first stage. The Future Circular proton-proton Collider (FCC) Study, currently under consideration by CERN, balances infrastructure and technology with physics potential for the future of particle physics \cite{fcc,fcc2}. The Chinese CEPC project (Circular Electron Positron Collider) is also proposed as a high-energy particle accelerator that would be built in China. The baseline design of the CEPC is based on a 50 km tunnel. Considering the same tunnel can house a Super Proton Proton Collider (SppC), its center of mass energy can range from 70-100 TeV. Therefore, benchmark SPPC physics studies are exploring the physics capabilities of proton-proton collisions at 100 TeV, and an integrated luminosity of 3 ab$^{-1}$ \cite{sppc}.

It is known that many experiments have observed neutrino oscillations produced by neutrino masses and mixing \cite{SNO:2001kpb, LSND:2001aii,KamLAND:2002uet,K2K:2006yov,os1, os2,os3,os4,os5}. The fact that neutrinos have a non-zero mass sparked interest in their electromagnetic properties. The electromagnetic properties of neutrinos are crucial because they relate directly to the fundamental principles of particle physics. For instance, the electromagnetic properties of neutrinos can be employed to differentiate between Dirac and Majorana neutrinos. Dirac neutrinos can possess both diagonal and off-diagonal magnetic dipole moments, while only off-diagonal moments are permissible for Majorana neutrinos. In the extension of the Standard Model (SM) with massive neutrinos, radiative corrections induce tiny couplings of $\nu\bar{\nu}\gamma$ and $\nu\bar{\nu}\gamma \gamma$ \cite{1,2,3,4,5,alex,alex1}. Even though minimal extension of the SM induces very small couplings, there are several models beyond the SM that predict relatively large couplings.
Hence, it is valuable to search for the electromagnetic properties of neutrinos in a model-independent way. Probing the electromagnetic structure of the neutrinos is important for understanding the physics beyond the SM and contributes to studies in astrophysics and cosmology. $\nu\bar{\nu}\gamma$ interactions are significant as they hold the potential to solve the mysterious solar neutrino puzzle, which could be the result of a large neutrino magnetic moment \cite{mom} or a resonant spin flip caused by Majorana neutrinos \cite{maj}. $\nu\bar{\nu}\gamma \gamma$ interactions could have significant impacts on a range of low- and high-energy reactions that are of astrophysical and cosmological interest \cite{cos}. For example, a high rate of photon annihilation into neutrino pairs may account for the observed cooling of stars via neutrino emission \cite{pon}. Furthermore, other intriguing processes that involve $\nu\bar{\nu}\gamma \gamma$ interactions include $\nu \gamma \to \nu \gamma$, $\nu\bar{\nu} \to \gamma \gamma$, and the neutrino double-radiative decay
$\nu_{i} \to \nu_{j} \gamma  \gamma$.

So far, experimentally derived limits on the neutrino magnetic moment from neutrino-electron scattering experiments with reactor neutrinos and solar neutrinos are at the order of $10^{-11}\mu_{B}$ \cite{m1,m2,m3,m4,m5,m6}. On the other hand, astrophysical observations provide us with more restrictive bounds. For example, the energy loss of astrophysical objects gives approximately an order of magnitude more restrictive bounds than those obtained from the reactor and solar neutrino probes \cite{m7,m8,m9,m10,m11,m12,Alok:2022pdn}.

A common method for exploring physics phenomena beyond the SM is by employing an Effective Field Theory approach. The way used to examine the non-standard neutrino-photon interactions with high-dimensional operators to obtain the effective vertices.
Non-standard $\nu\bar{\nu}\gamma \gamma$ coupling can be parametrized on dimension-seven operators at the lowest dimension.
The effective Lagrangian is defined as follows \cite{9,10,11,12,13,14}

\begin{eqnarray}
\label{eq.2}
\mathcal{L}=\frac{1}{4\Lambda^{3}}\bar{\nu}_{i}(\alpha_{R1}^{ij}P_{R}+\alpha_{L1}^{ij}P_{L})\nu_{j}{\widetilde F}_{\mu\nu}{F}^{\mu\nu}+\frac{1}{4\Lambda^{3}}\bar{\nu}_{i}(\alpha_{R2}^{ij}P_{R}+\alpha_{L2}^{ij}P_{L})\nu_{j}{F}_{\mu\nu}{F}^{\mu\nu}
\end{eqnarray}
where $\Lambda$ is the new physics scale, ${\widetilde F}_{\mu\nu}=\frac{1}{2}\epsilon_{\mu\nu\alpha\beta}{F}^{\alpha\beta}$, $P_{R}(L)=\frac{1}{2}(1\pm\gamma_{5})$, $\alpha_{Rk}^{ij}$ and $\alpha_{Lk}^{ij}$ are dimensionless coupling constants. In this study, we will examine the Dirac neutrino scenario and determine model-independent limits on the couplings $\alpha_1$ which is defined as $\alpha_1^2=\sum_{i,j}\left[|\alpha^{ij}_{R1}|^2+|\alpha^{ij}_{L1}|^2\right]$. 

The Lagrangian in Eq.(\ref{eq.2}) should also include terms involving a Z-boson due to the invariance of the theory in the UV under electroweak symmetry transformations. The four point vertex $Z \nu \bar{\nu} \gamma$ arises from the dimension-eight operators given in Refs. \cite{gul,tev} as;
\begin{equation} 
O^{8}_{1}={\mathrm{i}}
(\phi^{\dagger} \phi)\bar{\ell}^{a}_{L}
\tau^{i} \gamma^{\mu}
D^{\nu}\ell^{a}_{L}W^{i}_{ \mu \nu},
\label{eiop1} 
\end{equation}
\begin{equation} 
O^{8}_{2}= {\mathrm{i}}
(\phi^{\dagger} \phi)\bar{\ell}^{a}_{L}
\gamma^{\mu} D^{\nu} \ell^{a}_{L} B_{\mu \nu},
\label{eiop2} 
\end{equation}
\begin{equation}
O^{8}_{3}= {\mathrm{i}} (\phi^{\dagger}
D^{\mu} \phi)\bar{\ell}^{a}_{L} \gamma^{\nu}
\tau^{i} \ell^{a}_{L}W^{i}_{\mu \nu},
\label{eiop3}
\end{equation}
\begin{equation}
O^{8}_{4}= {\mathrm{i}} (\phi^{\dagger} D ^{\mu}
\phi)\bar{\ell}^{a}_{L}
\gamma^{\nu}
\ell^{a}_{L} B_{\mu \nu}. \label{eiop4}
\end{equation}
where $W^{i}_{\mu \nu}$ and $B_{\mu \nu}$, are denotes $\mathrm{SU(2)_{L}}$ and $ \mathrm{U(1)_{Y}}$
tensor field strength tensors 
respectively, as well as the  $\ell^{a}_{L}$ represents $\mathrm{SU(2)_{L}}$ left-handed lepton
doublet, $\tilde{\phi}=\mathrm{i} \tau^{2} \phi^{\ast}$ is the Higgs field, $\tau_{i}$ are the Pauli matrices and $D_{\mu}$ is covariant derivative. 
After spontaneous symmetry breaking, the operators
(\ref{eiop1})-(\ref{eiop4}) induce the following parametrization for $Z (q) \nu (p_2)\bar{\nu}(p_1) \gamma$(k) effective couplings,
\begin{equation} M^{(b)}_{\mu
\nu}=\frac{\epsilon_8}{v^2}\bar{u}(p_2)(1-\gamma_5)(k_{\mu}
\gamma_{\nu}-\slashed{k}g_{\mu \nu}) v(p_1),
\label{EcMb} \end{equation}
where $q=k+p_{2}$, $v$ is the SM vacuum expectation value and the
coefficients $\epsilon_{8}$  are expressed in terms of dimensionless coupling constants
$\alpha_{i}$ and the scale $\Lambda$:  
$\epsilon_{8}=\epsilon_{8}^{2} +
\epsilon_{8}^{3} + \epsilon_{8}^{4}$; with
$\epsilon^{i}_{8}=\alpha^{i}_{8}(v/\Lambda)^4$.
The Lagrangian in Eq.(\ref{eq.2}) is the most general dimension-seven Lagrangian describing $\nu\bar{\nu}\gamma \gamma$ coupling. However, as we mentioned above, the dimension-eight operators arising from $Z \nu \bar{\nu} \gamma$ coupling are suppressed by additional power of high inverse new physics scale $\Lambda$ and we shall ignore them in our study.


The upper bound on $\nu\bar{\nu}\gamma \gamma$ coupling was obtained from the LEP data through the $Z \to \nu\bar{\nu}\gamma \gamma$ decay as follows \cite{9}

\begin{eqnarray}
\label{eq.1}
[\frac{1 GeV}{\Lambda}]^{6}\sum_{i,j,k}(|\alpha_{Rk}^{ij}|^{2}+\alpha_{Lk}^{ij}|^{2})\leq 2.85\times10^{-9}.
\end{eqnarray}

Meanwhile, an analysis of the Primakoff effect on the conversion of $\nu_{\mu}N \to \nu_{\nu}N$ in the external Coulomb field of nucleus $N$ yields a bound that is two orders of magnitude more stringent than that obtained from the decay $Z \to \nu\bar{\nu}\gamma \gamma$ \cite{10}. Using the experimental bounds given an Eq.2 in the lifetime of the neutrino double radiative decay ($\nu_j\to\nu_i\gamma\gamma$) can help to make a connection with a typical neutrino mass. Considering the effective Lagrangian given in Eq.\ref{eq.2}, the expression of decay width of $\nu_j\to\nu_i\gamma\gamma$ without mass of $\nu_i$ is given as \cite{9}  
\begin{equation}
\label{eq3}
    \Gamma_{\nu_j\to\nu_i\gamma\gamma}=1.59\times10^{-9}(|\alpha_{Rk}^{ij}|^{2}+\alpha_{Lk}^{ij}|^{2})\Big[\frac{1 GeV}{\Lambda}\Big]^6\times\Big[\frac{m_{\nu_j}}{1MeV}\Big]^7 s^{-1}.
\end{equation}

Many phenomenological studies have been conducted on $\nu\bar{\nu}\gamma \gamma$ couplings at linear and hadron colliders. The best phenomenological obtained limits on $\alpha^2$ in the literature are $10^{-16}$ via exclusive $pp \rightarrow p\gamma^{*}\gamma^{*} p \rightarrow p \nu\bar{\nu}p$ process in  Ref.\cite {7} and $10^{-17}$ via the process $pp \rightarrow p\gamma^{*}\gamma^{*} p \rightarrow p \nu\bar{\nu}Zp$ (Here,$\gamma^{*}$ is  the Weizsacker-Williams photon)  at LHC energies. 
In Ref.\cite {8}, the sensitivities on $\nu\bar{\nu}\gamma\gamma$ couplings ($\alpha_1^2$ and $\alpha_2^2$) are obtained at the order of $10^{-17}$ and $10^{-18}$ via $\nu \bar{\nu}$ production in a $\gamma^{*} p$ collision at the LHC. The $e^{+}e^{-}$ linear colliders, including their $e\gamma$ and $\gamma\gamma$ operating modes, have been also explored through processes $e\gamma \to \nu\bar{\nu} e $ and $\gamma \gamma \to \nu\bar{\nu}$ (Here, $\gamma$ is  the laser photon) at the CLIC \cite{6}.

\section{Signal and background analysis}
\begin{figure}[h]
\centerline{\scalebox{0.8}{\includegraphics{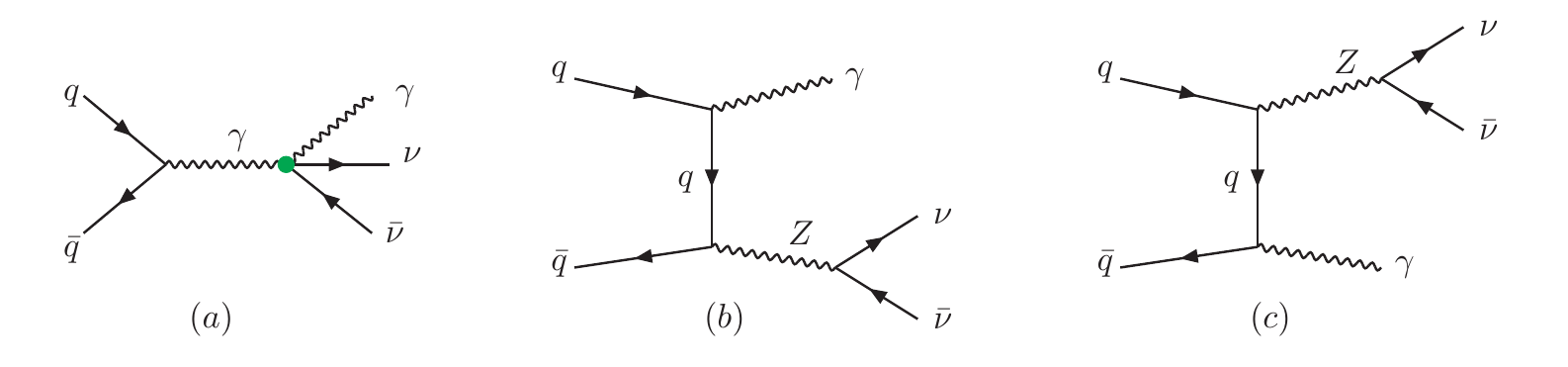}}}
\caption{Tree level Feynman diagrams for the process $pp\to \nu\bar{\nu}\gamma$ in the presence of non-standard $\nu\bar{\nu}\gamma \gamma$ coupling (green dot).}
\label{Fig.1}
\end{figure}

A detailed examination of the contributions of $\nu\bar{\nu}\gamma \gamma$ new physics vertices through the process $pp\to \nu\bar{\nu}\gamma$ and the relevant SM backgrounds is carried out at the HL-LHC with $\sqrt s=$14 TeV and the FCC-hh/SppC with $\sqrt s=$100 TeV. The Feynman diagrams for the signal process $pp \to \nu\bar{\nu}\gamma$ at the tree level are displayed in Fig. \ref{Fig.1}. The first diagram accounts for the anomalous couplings, while the rest represents the SM contributions. The calculated analytical expressions for the polarization summed amplitude square of the $pp \to \nu\bar{\nu}\gamma$ process using an effective Lagrangian (Eq.~(\ref{eq.2})) are given as follows
\begin{eqnarray}
\langle|M|^2\rangle&=&\sum_{i,j}\langle|M_a|^2\rangle+\langle|M_b+M_c|^2\rangle
\end{eqnarray}
where $M_a$ is the contribution when the effective interaction $\nu\bar{\nu}\gamma \gamma$  is taken into account, while $M_b$ and $M_c$ are the SM contributions. The interference terms between $M_b$ and $M_c$ contribute to the total amplitude, while the interference terms between $M_a$ and $M_b$ and between $M_a$ and $M_c$ do not contribute. Since the momentum dependence of the new physics contribution from the $\nu\bar{\nu}\gamma \gamma$ vertices in the amplitude square is the same for both $\alpha_{L1(R1)}^{ij}$ and $\alpha_{L2(R2)}^{ij}$, we will focus on the analysis of $\alpha_{L1(R1)}^{ij}$ couplings. 
For further detailed numerical analysis, the effective Lagrangian given in Eq.(\ref{eq.2}) is implemented in the {\sc FeynRules} \cite{fey} package to generate a {\sc Universal FeynRules Output} (UFO) module \cite{ufo} which is inserted to {\sc MadGraph5\_aMC$@$NLO v3\_1\_1}\cite{mg5}. 
 
\begin{figure}[h]
\centerline{\scalebox{0.7}{\includegraphics{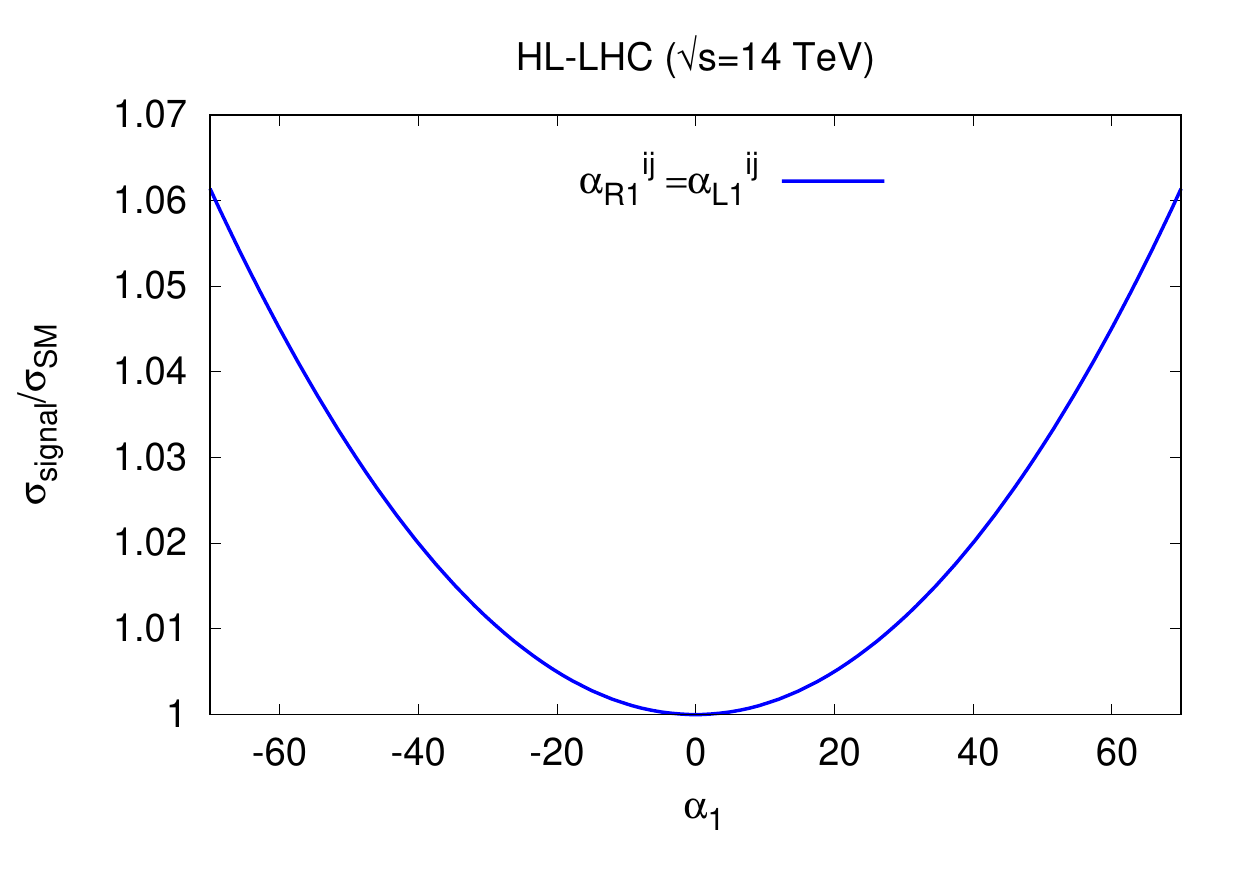}}\scalebox{0.7}{\includegraphics{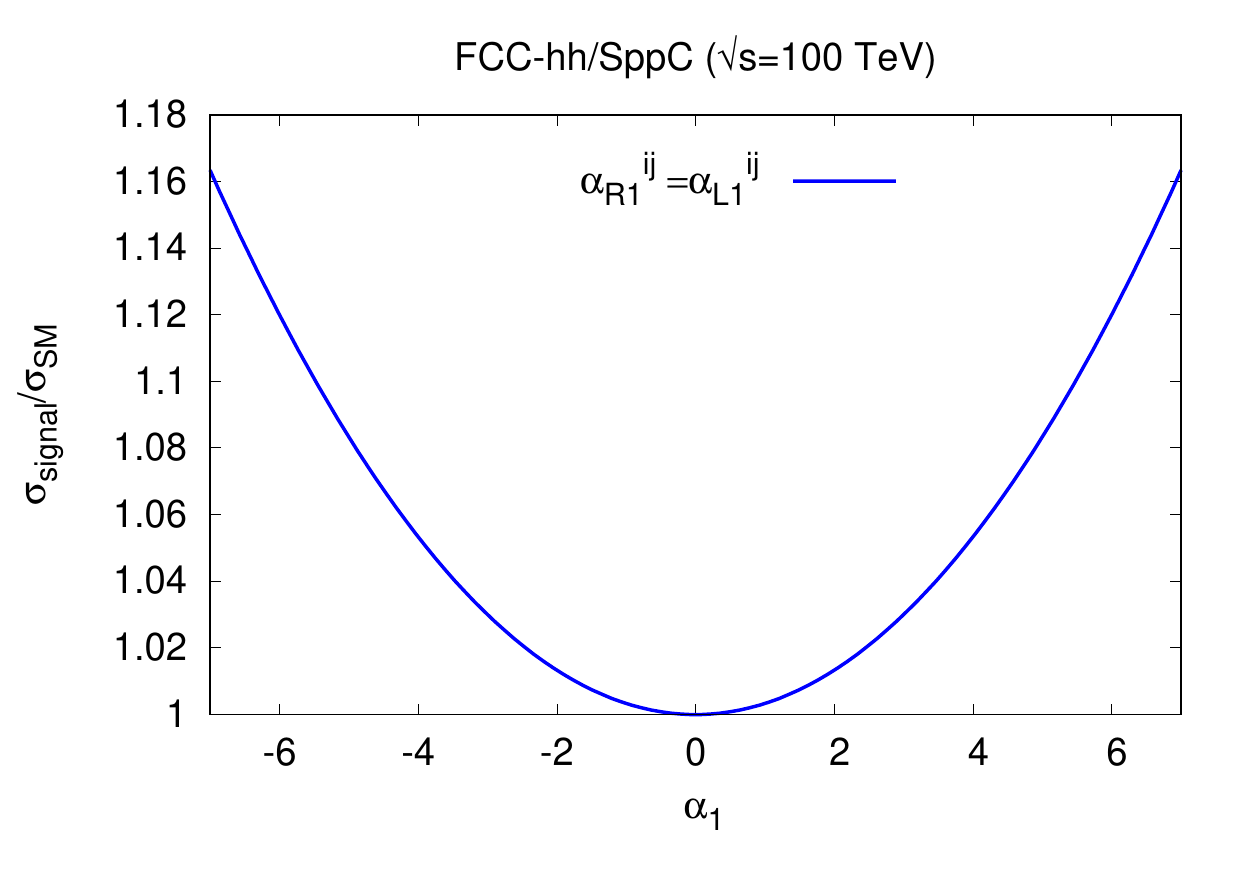}}}
\caption{The signal-to-SM ratio of total cross-sections for the process $pp\to \nu\bar{\nu}\gamma$ as a function of the anomalous $\alpha_1$ coupling  at the HL-LHC and the FCC-hh.}
\label{Fig.2}
\end{figure}

In this paper, we have set $\Lambda$= 1 TeV as a reference value for our calculations. Fig.~\ref{Fig.2} shows the signal-to-SM ratio of the cross-section for the $pp \to \nu\bar{\nu}\gamma$ process as a function of the $\alpha_1$ coupling for the case ($\alpha_{L1}^{ij} = \alpha_{R1}^{ij}$). The left panel displays the results for the HL-LHC, while the right panel shows the results for FCC-hh/SppC, with generator-level cuts of $p_T^{\gamma} >$ 10 GeV and $|\eta^{\gamma}| < 2.5$.
The deviation from the SM cross-section value begins around $\mathcal{O}$(1) for the HL-LHC and $\mathcal{O}$(10) for the FCC-hh/SppC. Considering experimental distinguishability on the cross section due to systematics, we have chosen the values of $\alpha_1$ that cover the cross sections that do differ from SM by 1\% to 10\% for HL-LHC and 1\%- 15\% for FCC-hh for detailed analysis.

Monte Carlo (MC) event samples for both the signal and relevant SM background processes are generated using {\sc MadGraph5\_aMC$@$NLO v3\_1\_1} with NNPDF23LO PDF set \cite{Ball:2012cx}, and then passed through {\sc Pythia 8.2} \cite{pyt} to include the initial and final parton showering and fragmentation. The detector response is simulated using {\sc Delphes 3.4.2} \cite{del}, utilizing the HL-LHC and the FCC-hh configuration cards. Jets are reconstructed by using clustered energy deposits with {\sc FastJet 3.3.2} \cite{Cacciari:2011ma} using anti-kt algorithm \cite{Cacciari:2008gp}. In order to perform detailed signal and background analysis, we generate 600k event samples for all relevant background processes and signals, for which six different benchmark points range from $\alpha_{L1}^{ij}=\alpha_{R1}^{ij}$= 7.0  (0.7) to $\alpha_{L1}^{ij}=\alpha_{R1}^{ij}$=70.0 (7.0) for HL-LHC (FCC-hh/SppC).

The final state in the signal process, $pp \rightarrow \nu\bar\nu\gamma$, consists of a photon and missing transverse energy ($\slashed {E}_T$). Therefore, we consider relevant SM background processes at hadron colliders with similar or identical final state topologies, including $\nu \bar{\nu} \gamma $, $ W \gamma $, $\gamma j$, and $Z (l l) \gamma $. We chose these SM backgrounds for several reasons, including that the process with the same final state as the signal process ($\nu \bar{\nu} \gamma $) is considered the dominant background, with events involving real $\slashed {E}_T$ from neutrinos. The $W\gamma$ background process, which includes the leptonic decay of the $W$, can be defined as an electroweak process where no leptons are detected, potentially satisfying the condition of $\slashed {E}_T$ and at least one photon in the final state. The $\gamma j$ background process contains events involving prompt photons and mismeasured jet momenta, leading to missing transverse momentum, and therefore has the potential to share the same final state topology as the signal process. Another relevant SM background is $Z(ll)\gamma$ production, which involves events with incorrectly measured lepton momenta where $l=e,\mu,\tau$. 

The fake photon-lepton ($f_{e\to\gamma}$) and jet-photon ($f_{j\to\gamma}$) rates can be parametrized depending on the transverse momentum ($p_T^{\gamma}$) and pseudo-rapidity ($\eta^{\gamma}$ of photon. At the LHC, these ratios range from 0.6\% to 2.7\% for $f_{e\to\gamma}$ and can be formulated as $0.0007\cdot e^{-p_T[GeV ]/187}$ for $f_{j\to\gamma}$ \cite{ATLAS:2016ukn}. In our analysis, we applied a fixed jet-to-photon fake rate of $0.1\%$ in the $\gamma j$ process and a lepton-to-photon fake rate of 2\% in the $W\gamma$ and $Z(ll)\gamma$ processes in order to model the smooth background for the HL-LHC. Even though the future hadron-hadron collider detectors are expected to have a granularity that is 2-4 times better than the current HL-LHC detectors, we considered the same fake rate as for the HL-LHC detectors for the FCC-hh/SppC options \cite{Mangano:2020sao}.

\begin{figure}[htb!]
\centerline{\scalebox{0.4}{\includegraphics{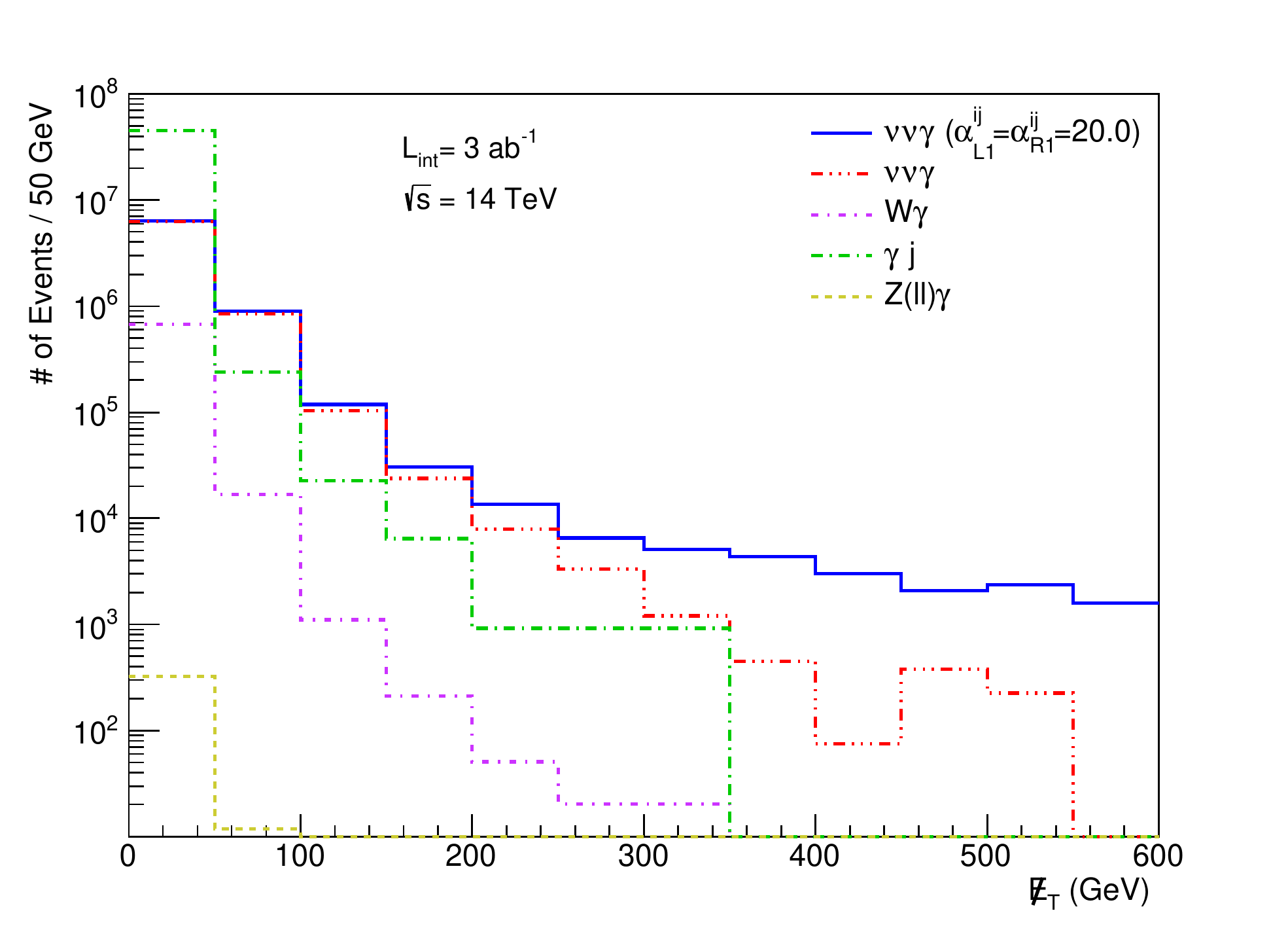}}\scalebox{0.4}{\includegraphics{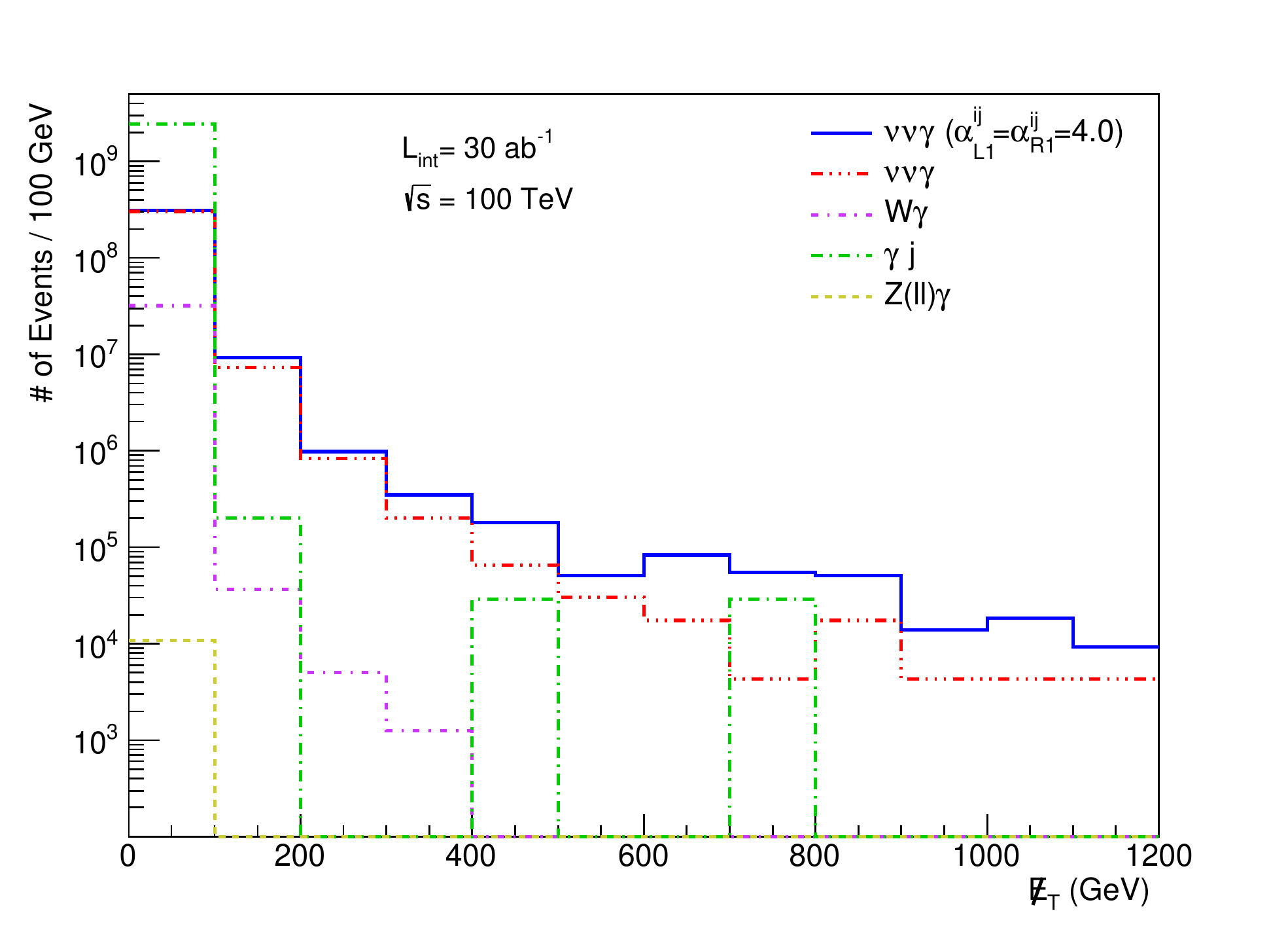}}}
\caption{The normalized distribution of $\slashed {E}_T$ for the process $pp\to \nu\bar{\nu}\gamma$ signal and backgrounds at the HL-LHC (on the left) and the FCC-hh/SppC (on the right). The dotted lines show the background processes and the solid lines represent the signal process.}
\label{Fig.3}
\end{figure}

\begin{figure}[t]
\centerline{\scalebox{0.4}{\includegraphics{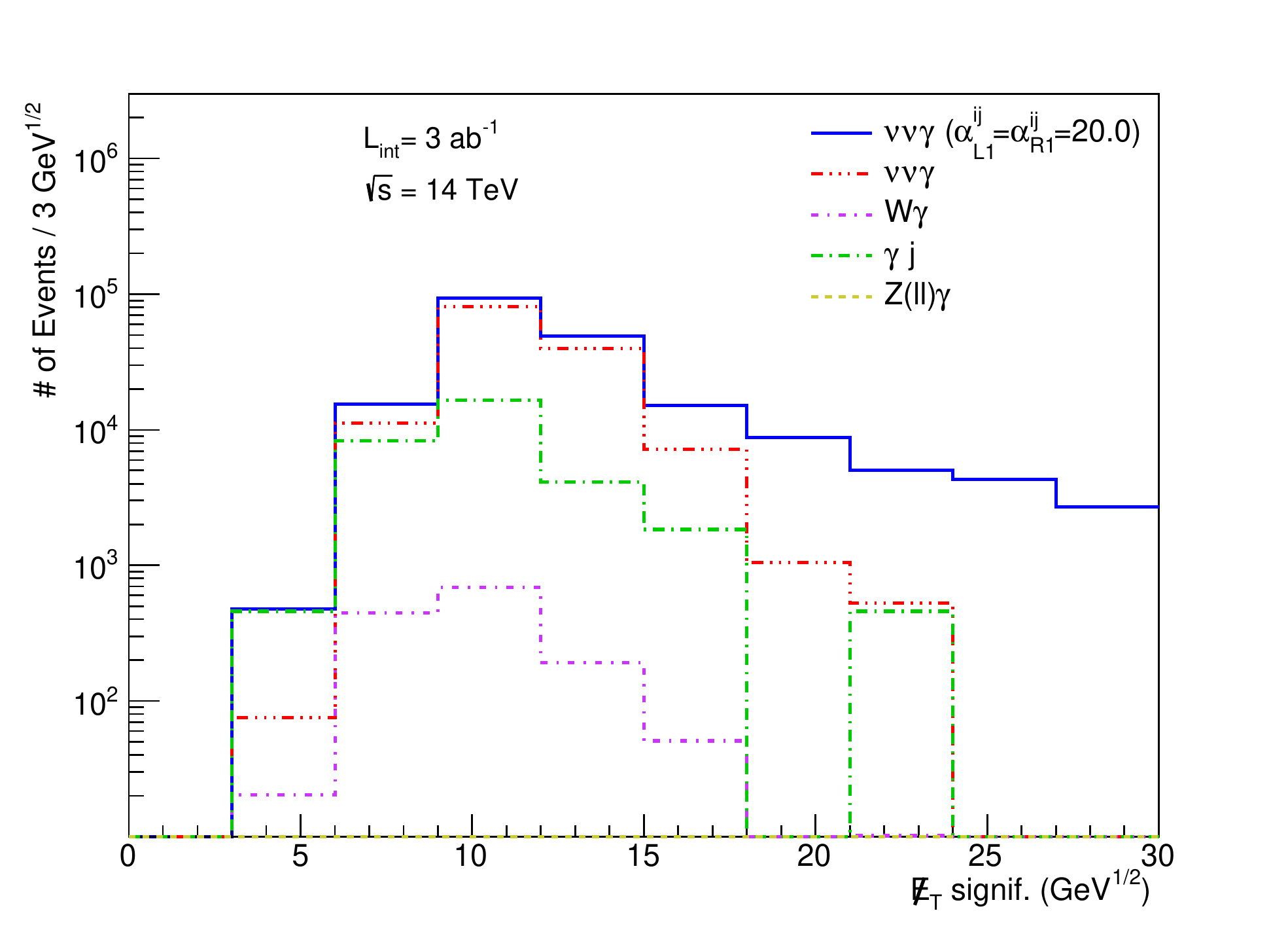}}\scalebox{0.4}{\includegraphics{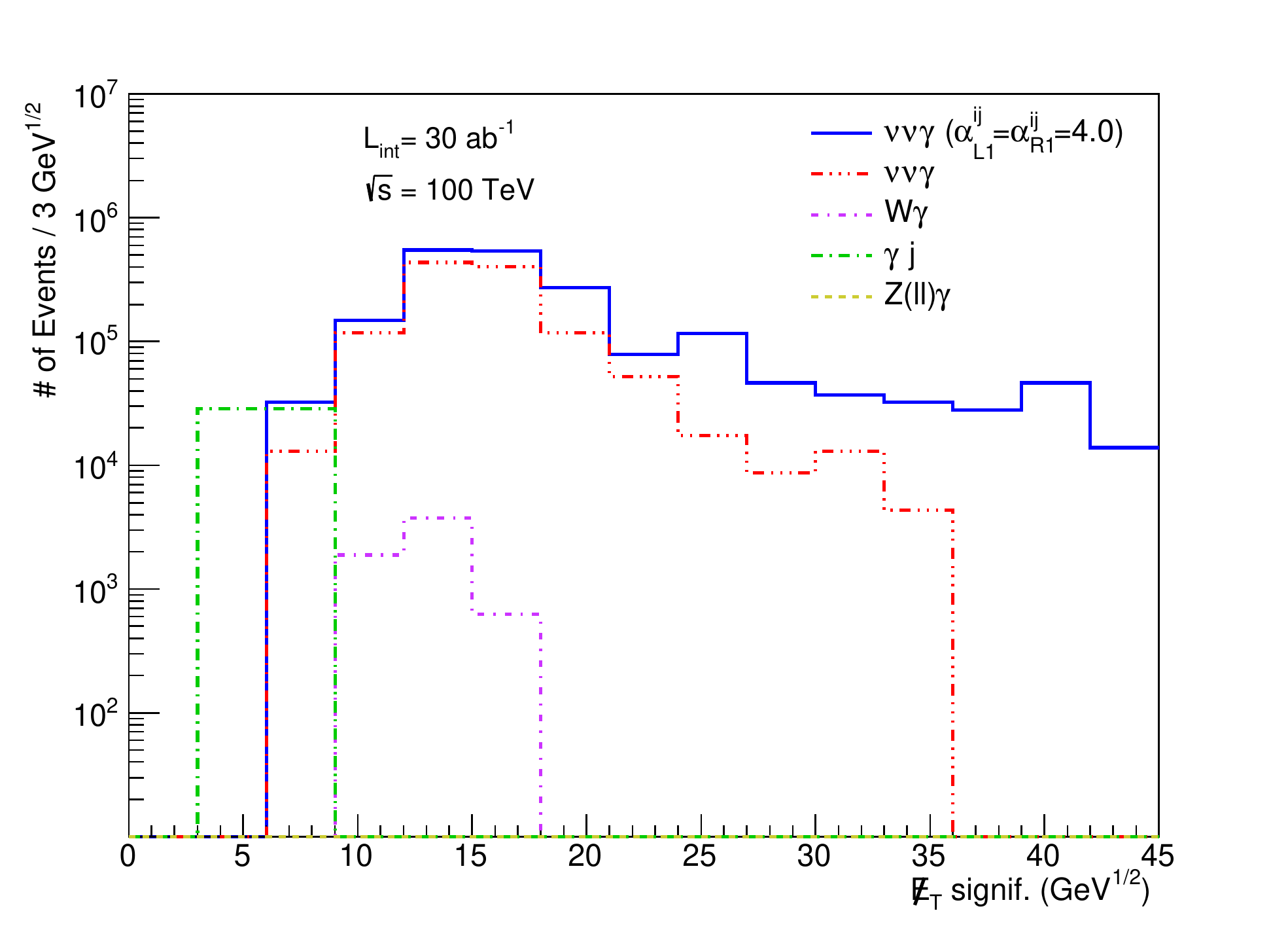}}}
\caption{The normalized distributions of $\slashed {E}_T$ signif. for the process $pp\to \nu\bar{\nu}\gamma$ signal and backgrounds at the HL-LHC (on the left) and the FCC-hh/SppC (on the right). The dotted lines show the background processes and the solid lines represent the signal process.}
\label{Fig.4}
\end{figure}

The criteria for event selection are designed not only to achieve high sensitivity to any deviations from the expected non-standard $\nu\bar{\nu}\gamma\gamma$ couplings but also to achieve both high signal efficiency and effective rejection of the relevant SM backgrounds. Therefore, events are pre-selected based on the final state topology of the signal process, requiring at least one photon with a non-zero $\slashed {E}_T$ and no leptons. The normalized distributions of $\slashed {E}_T$ for the signal process $pp\to \nu\bar{\nu}\gamma$ and background processes at the HL-LHC with $L_{int}=3$ ab$^{-1}$ and $\sqrt{s}=14$ TeV (for $\alpha_{L1}^{ij}=\alpha_{R1}^{ij}=20.0$) and at the FCC-hh/SppC with $L_{int}=30$ ab$^{-1}$ and $\sqrt{s}=100$ TeV (for $\alpha_{L1}^{ij}=\alpha_{R1}^{ij}=4.0$) are shown in Fig.~\ref{Fig.3}. To reduce contamination by fake high-energy neutrinos ($\slashed {E}_T$) which is mainly due to inaccurate measurements of jet momentum in the $\gamma j$ background process, we require selected events to have $\slashed {E}_T>$ 150 (200 GeV) for the HL-LHC (the FCC-hh/SppC). We introduce a missing transverse energy significance ($\slashed {E}_T$ signif.) variable to help distinguish real missing transverse energy due to undetected particles from fake missing transverse energy due to object mis-reconstruction, finite detector resolution, or detector noise \cite{ATLAS:2018nci}. $\slashed {E}_T$ significance is defined as $\slashed {E}_T / \sqrt{\sum p_T^j + p_T^{\gamma}}$. In Fig.~\ref{Fig.4}, we show the normalized distributions of $\slashed {E}_T$ significance for the signal and background processes at the HL-LHC and the FCC-hh/SppC. By applying a cut on $\slashed {E}_T$ signif.$>10.5$ GeV$^{1/2}$ for both collider options, it is possible to further suppress background contributions with fake $\slashed {E}_T$.

\begin{figure}[t]
\centerline{\scalebox{0.4}{\includegraphics{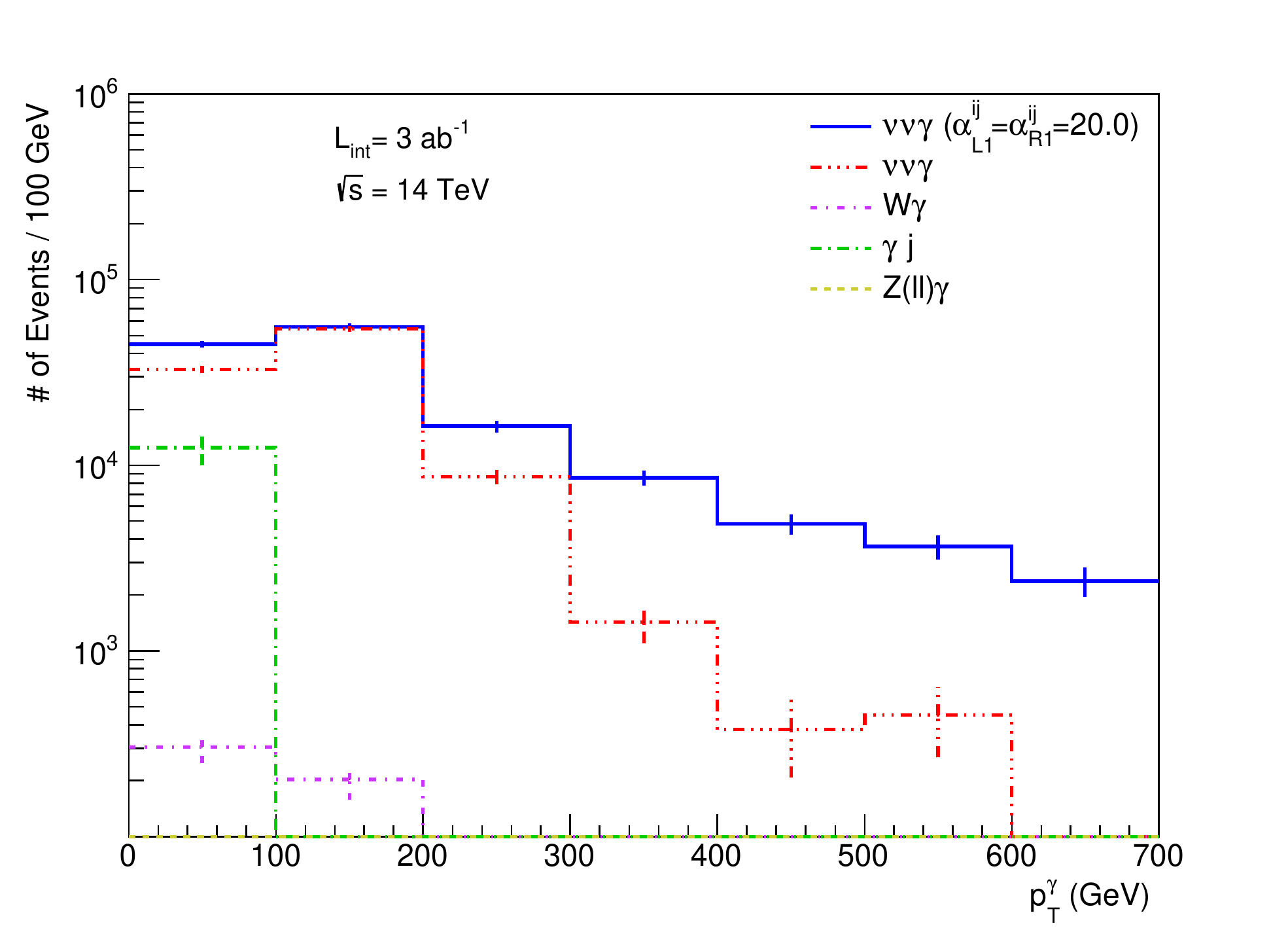}}\scalebox{0.4}{\includegraphics{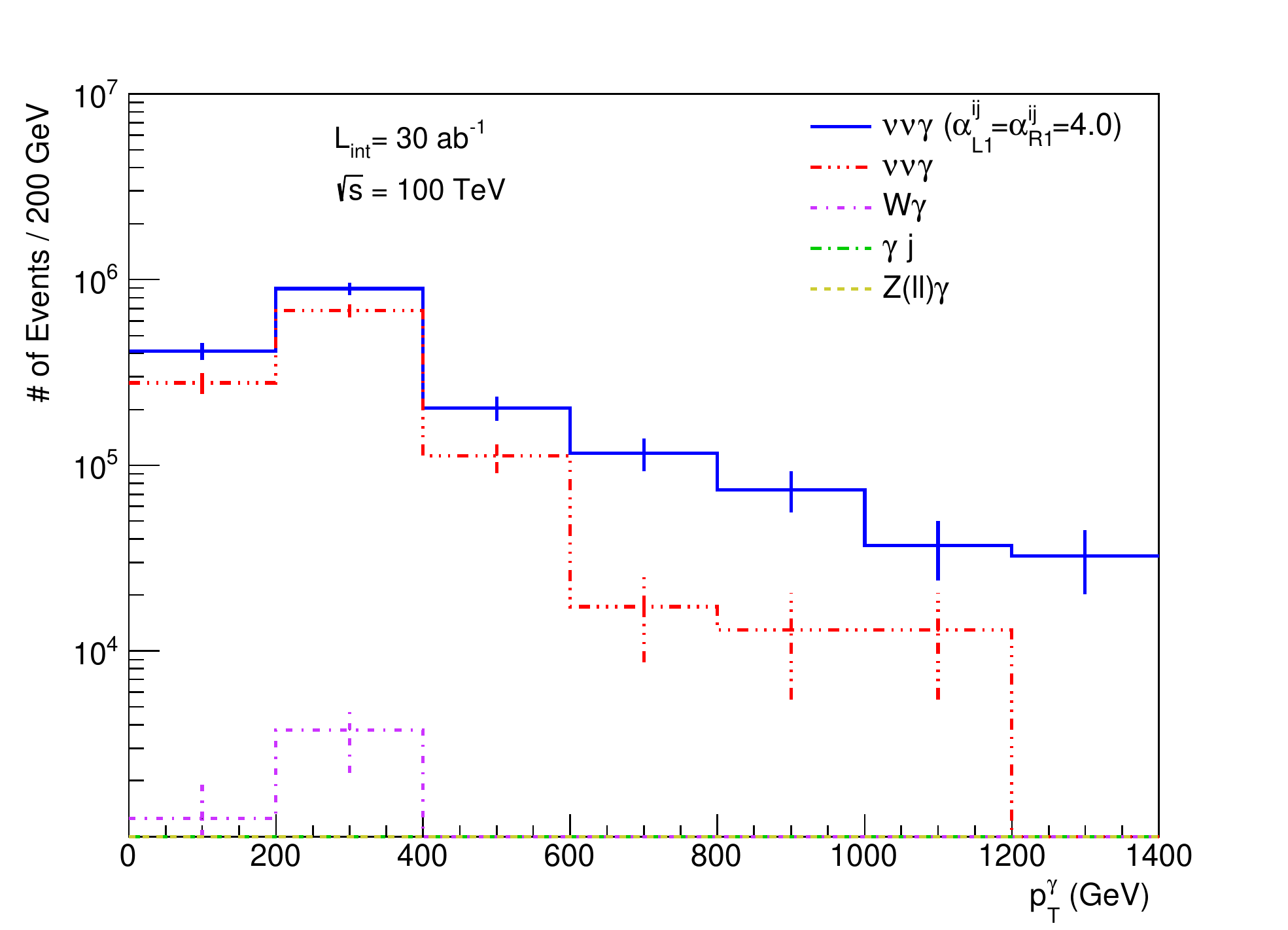}}}
\caption{The normalized distributions of $p_{T}^{\gamma}$ for the process $pp\to \nu\bar{\nu}\gamma$ signal and backgrounds at the HL-LHC (on the left) and the FCC-hh/SppC (on the right). The dotted lines show the background processes and the solid lines represent the signal process.}
\label{Fig.5}
\end{figure}

Selecting $\nu\bar{\nu}\gamma$ events in the high $p_{T}^{\gamma}$ region brings new physics due to non-standard neutrino interactions to the forefront. Fig.~\ref{Fig.5} displays the normalized distribution of $p_{T}^{\gamma}$ for the $pp \to \nu\bar{\nu}\gamma$ signal and relevant backgrounds at the HL-LHC and the FCC-hh/SppC. As seen in Fig.~\ref{Fig.5}, new physics parameters defined by high-dimensional operators may impact the transverse momentum of the photon, particularly at high $p_{T}^{\gamma}$ regions, making it easier to distinguish the signal from relevant backgrounds. Hence, we apply cuts on $p_{T}^{\gamma}> 250~(400)$ GeV, where the signal deviates from the SM backgrounds, for further analysis at the HL-LHC (the FCC-hh/SppC). The cuts used in the analysis are summarized in Table \ref{cut}. The number of signal and relevant background events normalized to the corresponding integrated luminosity after each cut with the statistical uncertainties are given in Table \ref{nof_HLLHC} (\ref{nof_FCChh}) for HL-LHC (FCC-hh/SppC), where effects of the cuts can be seen. As can be seen from Table \ref{nof_HLLHC} (\ref{nof_FCChh}), the statistical uncertainty in the $W\gamma$, $\gamma j$, and $Z(ll)\gamma$ background processes is about four times larger than the uncertainty in the signal and $\nu\nu\gamma$ background after the Pre-selection and Cut-1. This is also seen as statistical fluctuations in the region $\slashed {E}_T>$ 200 (400) GeV in Fig.\ref{Fig.3} which is plotted after Pre-selection cut, and in the region $\slashed {E}_T$ signif. $> 10.5$ in Fig.\ref{Fig.4}, which is plotted after Cut-1. The statistical fluctuations in these regions are due to the limited number of MC events. Table \ref{nof_HLLHC} (\ref{nof_FCChh}) also shows that the statistical uncertainty in the number of events obtained after Cut-3 is very low for both the $\nu\nu\gamma$  background and the signal. The statistical fluctuations are absent in both the $\nu\nu\gamma$ background and the signal distributions in the region $p_T> 250 (400)$ GeV of Fig. \ref{Fig.5}. Since the main contribution to the total number of background events comes from the $\nu\nu\gamma$ background and there is almost no contribution from the other backgrounds, these fluctuations will not affect our study to obtain the limits on the couplings of the non-standard $\nu\bar{\nu}\gamma \gamma$ vertex.

\begin{table}[h]
\caption{Event selection and kinematic cuts used for signal and background events at the HL-LHC (the FCC-hh/SppC) analysis \label{cut}.}
\begin{ruledtabular}
\begin{tabular}{lccc}
 Pre-selection & $N_{\gamma} = 1$, $\slashed {E}_T>$ 0  and $N_l = 0$ $|\eta^{\gamma}|<2.5$ and $p_{T}^{\gamma} >$ 10   \\
 Cut-1 &$\slashed {E}_T > 100 (200)$ GeV \\
 Cut-2 &$\slashed {E}_T$ signif. $> 10.5$ GeV$^{1/2} $ \\
Cut-3 & $p_{T}^{\gamma} > 250 ~(400)$  GeV
\end{tabular}
\end{ruledtabular}
\end{table}

\begin{table}[h]
\caption{Number of event for signal (for $|\alpha^{ij}_{R1}|=|\alpha^{ij}_{L1}|=20$) and background with statistical uncertainty after each applied cuts at the HL-LHC  analysis. \label{nof_HLLHC}}
\begin{ruledtabular}
\begin{tabular}{lcccc}
Cuts&  Pre-selection &Cut-1&Cut-2&Cut-3\\\hline
Signal&$7.4018\cdot10^6$ $\pm$ 2721& 198936 $\pm$ 446&145548 $\pm$ 382 & 34241 $\pm$ 185\\  
$\nu \bar{\nu} \gamma $& $7.24592\cdot10^6$ $\pm$ 2692 & 141324 $\pm$ 376 & 98106 $\pm$ 313& 5496$\pm$ 74 \\
$ W \gamma $& 689355 $\pm$ 830 &14010 $\pm$ 38 & 538 $\pm$ 23 & 10 $\pm$ 3\\
$\gamma j$& $4.54179\cdot10^6$ $\pm$ 6739&31730 $\pm$178 &12416 $\pm$ 111 &0 \\
$Z (l l) \gamma$ & 335 $\pm$ 18 & 1.17 $\pm$ 1.08 & 0& 0\\
\end{tabular}
\end{ruledtabular}
\end{table}

\begin{table}[h]
\caption{Number of events for signal  (for $|\alpha^{ij}_{R1}|=|\alpha^{ij}_{L1}|=4$) and background with statistical uncertainty after each applied cuts at the FCC-hh/SppC analysis.\label{nof_FCChh}}
\begin{ruledtabular}
\begin{tabular}{lcccc}
Cuts&  Pre-selection &Cut-1&Cut-2&Cut-3\\\hline
Signal&$3.24743\cdot10^8$ $\pm$ 1821& $2.04637\cdot10^6$ $\pm$ 1430 & $1.97692\cdot10^6$ $\pm$ 1406 & 671319 $\pm$ 819\\  
$\nu \bar{\nu} \gamma $& $3.10327\cdot10^8$ $\pm$ 17616 & $1.17914\cdot10^6$ $\pm$ 1086  & $1.11845\cdot10^6$ $\pm$ 1057 & 160398$\pm$ 400 \\
$ W \gamma $&$3.21456\cdot10^7$ $\pm$ 5670 &6243 $\pm$ 79 & 4994 $\pm$ 71 & 0\\
$\gamma j$& $2.46105\cdot10^9$ $\pm$ 49609 & 57529 $\pm$290 &0  &0 \\
$Z (l l) \gamma$ & 10866 $\pm$ 104 & 4.31 $\pm$ 2.08 & 2.59 $\pm$ 1.61 & 0\\
\end{tabular}
\end{ruledtabular}
\end{table}

\begin{figure}[t]
\centerline{\scalebox{0.4}{\includegraphics{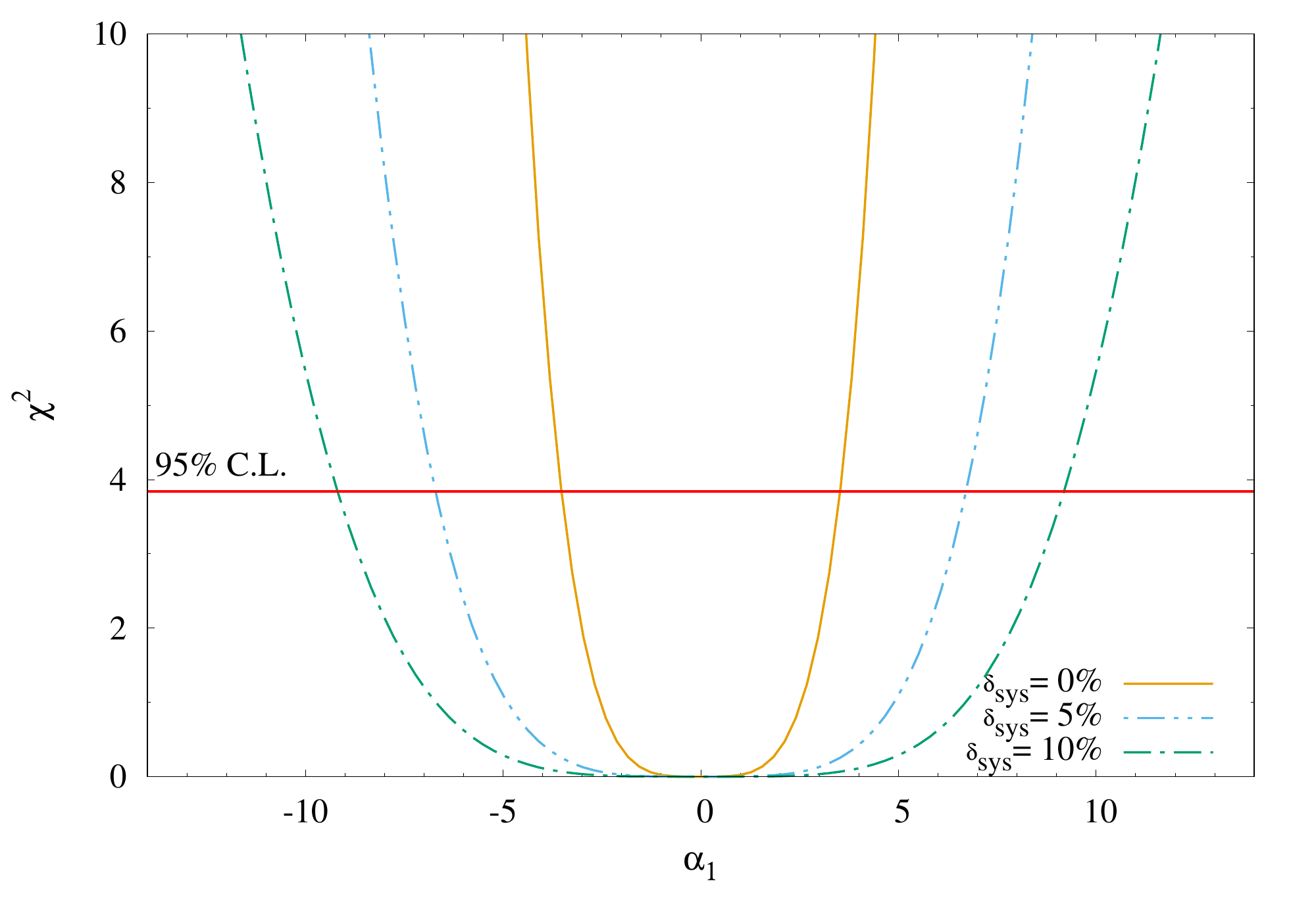}}\scalebox{0.4}{\includegraphics{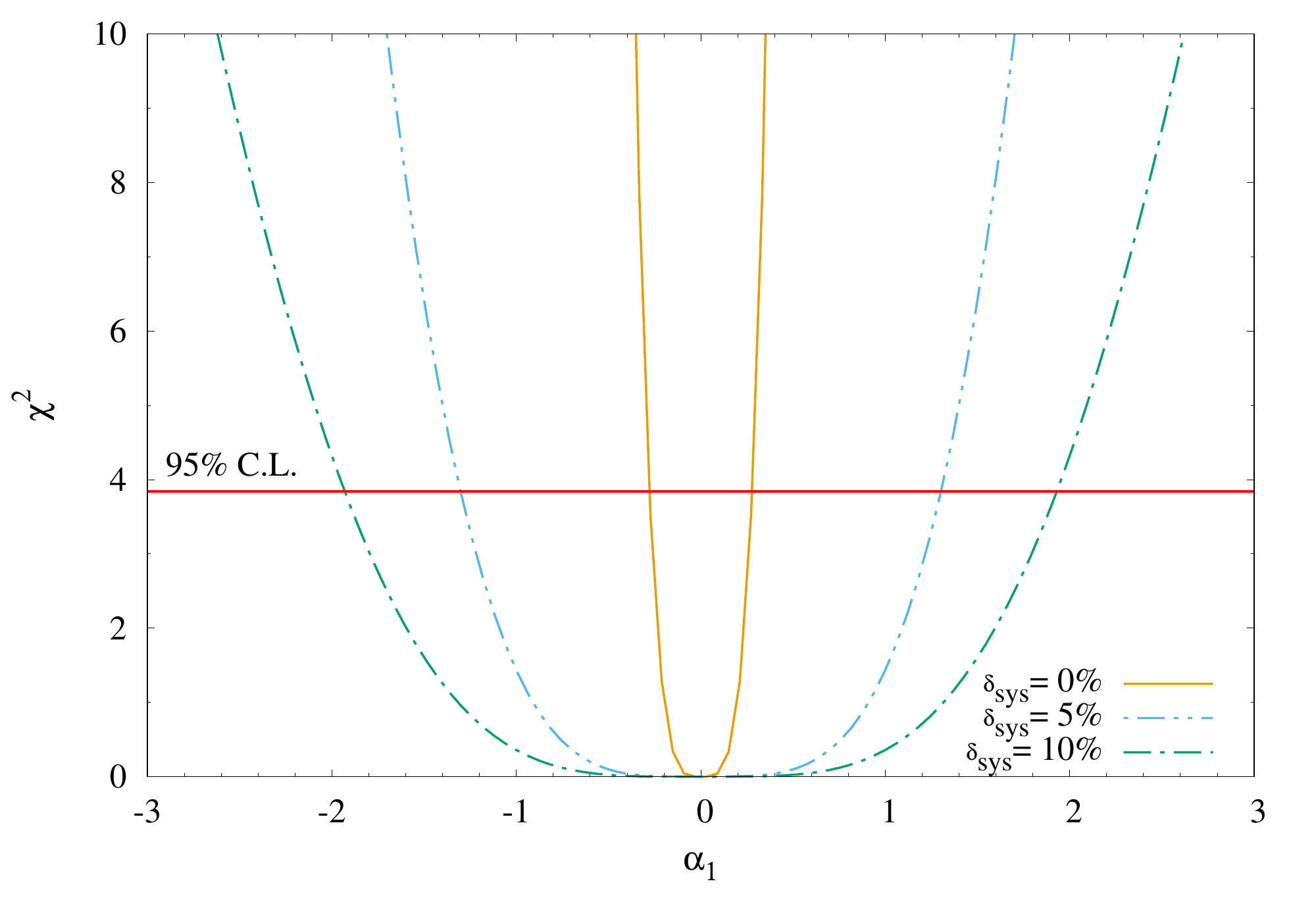}}}
\caption{The $\chi^2$ as function of $\alpha_1$ coupling for the process $pp\to \nu\bar{\nu}\gamma$ with selection strategy described in the text is used for both the HL-LHC (on the left) and the FCC-hh/SppC options (on the right).}
\label{Fig.6}
\end{figure}

\section{Sensitivity on the anomalous coupling}

To obtain sensitivity to the anomalous couplings through the process $pp \to \nu\bar{\nu}\gamma$, we employ the $\chi^{2}$ method. The $\chi^{2}$ function with and without a systematic error is defined as follows

\begin{eqnarray}
\chi^{2} =\sum_i^{n_{bins}}\left(\frac{N_{i}^{NP}-N_{i}^{B}}{N_{i}^{NP}\Delta_i}\right)^{2}
\end{eqnarray}
where $N_{i}^{NP}$ and $N_{i}^{B}$ are the total number of events in the existence of effective couplings and relevant SM backgrounds obtained after applying Cut-3 integrating the transverse momentum distribution of photon, respectively; $\Delta_i=\sqrt{\delta_{sys}^2+\frac{1}{N_i^{NP}}}$ is the combined systematic ($\delta_{sys}$) and statistical uncertainties in each bin.

In order to obtain a continuous prediction for the non-standard neutrino couplings, a quadratic fit is performed to the number of events for each coupling obtained by applying the cuts defined in Table \ref{cut}. All figures (Figs. ~\ref{Fig.2}-~\ref{Fig.5}) and the number of events used in the analysis are normalized to the cross-section of each process times the integrated luminosity, $L_{int}$ = 3 ab$^{-1}$ for the HL-LHC and $L_{int}$ = 30 ab$^{-1}$ for the FCC-hh/SppC.
When presenting a realistic physics potential of the HL-LHC and the FCC-hh/SppC for the process $pp\to \nu\bar{\nu}\gamma$, multiple assumptions can be made about the evolution of sources of uncertainty. Since the primary objective of this study is to explore the overall impact of systematic uncertainty on the limit of non-standard dimension-seven neutrino-two photon couplings rather than delving into the sources of such uncertainty, we present our obtained limits without and with systematic uncertainties 5\% and 10\% for the optimistic and realistic scenarios, respectively, for each collider. Fig. \ref{Fig.6} shows 95\% C.L. line as well as the $\chi^2$ as functions of the $\alpha_1$ coupling, considering without and with systematic uncertainties for the HL-LHC (on the right) at $L_{int}$ = 3 ab$^{-1}$ and the FCC-hh/SppC (on the left) at $L_{int}$ = 30 ab$^{-1}$.
From these figures, limits on dimension-seven combined neutrino-two photon coupling $\alpha_1^2=\sum_{i,j}\left[|\alpha^{ij}_{R1}|^2+|\alpha^{ij}_{L1}|^2\right]$ can be inferred from the intersection of curves with the horizontal 95\% C.L. red line and are listed in Table \ref{tab3}. We can see from this table that our obtained limits on $\alpha_1^2$ are approximately at the order of $10^{1}$ and $10^{-2}$ for the HL-LHC at $L_{int}$ = 3 ab$^{-1}$ and the FCC-hh/SppC at $L_{int}$ = 30 ab$^{-1}$ without systematic uncertainties. The best phenomenological obtained limits on $\alpha^2$ in the literature are $10^{2}$ via exclusive $pp \rightarrow p\gamma^{*}\gamma^{*} p \rightarrow p \nu\bar{\nu}p$ process in  Ref.\cite {7} and $10^{1}$ via the process $pp \rightarrow p\gamma^{*}\gamma^{*} p \rightarrow p \nu\bar{\nu}Zp$  at LHC energies when $\Lambda$ is rescaled to 1 TeV. In Ref.\cite {8}, the sensitivities on $\nu\bar{\nu}\gamma\gamma$ couplings ($\alpha_1^2$ and $\alpha_2^2$) are obtained at the order of $10^{1}$ and $10^{0}$  with rescaling $\Lambda$ via $\nu \bar{\nu}$ production in a $\gamma^{*} p$ collision at the LHC.

Finally, our obtained bounds on the neutrino-two-photon interaction taking into account $\Lambda$ =1 TeV also allow us to constrain the width of the $\nu_j\to\nu_i\gamma\gamma$  decay given in Eq.(\ref{eq3}), which can be translated into a constraint on the lifetime of $\nu_j$ as long as $m_{\nu_j}$ is in the range of a few tenths of keV.
\begin{table}
\caption{The obtained limits at $95\%$ C.L. on $\alpha_1^2=\sum_{i,j}\left[|\alpha^{ij}_{R1}|^2+|\alpha^{ij}_{L1}|^2\right]$ couplings ($\Lambda$ is set to 1 TeV) for the process $pp\to \nu\bar{\nu}\gamma$ at the HL-LHC and the FCC-hh/SppC with various values of the luminosity.}
\label{tab3}
\begin{ruledtabular}
\begin{tabular}{lcc}
& HL-LHC($L_{int}$ = 3 ab$^{-1}$)& FCC-hh/SppC($L_{int}$ = 30 ab$^{-1}$) \\
\hline
 $\delta_{sys}=0$ &$1.26\times 10^{1}$ &$7.76\times 10^{-2}$  \\

$\delta_{sys}=5\%$ &$4.50\times 10^{1}$&$1.69\times 10^{0}$    \\

$\delta_{sys}=10\%$ &$8.44\times 10^{1}$&$3.73\times 10^{0}$  \\

\end{tabular}
\end{ruledtabular}
\end{table}

\section{Conclusions}

The exploration of neutrino properties and interactions, both theoretically and experimentally, is one of the most active fields of research in current particle physics. This research provides valuable insights into the physics of the SM and offers a significant opportunity to explore physics beyond the SM. Thanks to numerous experiments, detecting neutrino oscillations leads to non-zero neutrino mass, which requires the SM to be extended. In many of these extensions, neutrinos can also possess electromagnetic properties due to quantum loop effects, leading to direct interactions with electromagnetic fields and charged particles. Therefore, the investigation of neutrino electromagnetic interactions is a powerful tool in the search for the fundamental theory beyond the SM. 

In this paper, we investigate the potential of the process $pp \rightarrow \nu\bar{\nu}\gamma$ at the HL-LHC and the FCC-hh/SppC to examine $\nu\bar{\nu}\gamma\gamma$ couplings. Comparing our obtained 95\% C.L. limits on $\alpha_1^2$ given in Table \ref{tab3}  with only available experimental LEP bound \cite{9} and obtained phenomenological limits in the literature \cite{7,8}, sensitivity on $\nu\bar{\nu}\gamma\gamma$ couplings improves at the two (one) orders with those reported in Ref.\cite {7} (Ref.\cite {8}) while a factor of $10^{10}$ with respect to the LEP limit. Even including 10\% systematic error, our results are much better than current available bounds in the literature. 

Finally, the process $pp \rightarrow \nu\bar{\nu}\gamma$ is sensitive to the anomalous $\nu\bar{\nu}\gamma\gamma$ couplings and worth investigating not only in the current LHC experimental data but also in the future hadron-hadron colliders with the expected integrated luminosity since it promises a significant improvement upon the LEP results.

\end{document}